\documentclass{ws-ijmpa}

\usepackage{graphicx}

\newcommand{\srt}{$\sqrt{s_{_{\rm NN}}}$}

\newcommand{\pt}{$p_T$ }
\newcommand{\pto}{$p_T$}
\newcommand{\GeVc}{GeV/$c$ }
\newcommand{\etal}{{\it et al.}}
 
\begin{document}
 
%
%
\def\Journal#1#2#3#4{{#1} {\bf #2} (#4) #3}
\def\NCA{\em Nuovo Cimento}
\def\NIM{\em Nucl. Instr. Meth.}
\def\NIMA{{\em Nucl. Instr. Meth.} A}
\def\NPB{{\em Nucl. Phys.} B}
\def\NPA{{\em Nucl. Phys.} A}
\def\PLB{{\em Phys. Lett.}  B}
\def\PRL{{\em Phys. Rev. Lett.}}
\def\PRC{{\em Phys. Rev.} C}
\def\PRD{{\em Phys. Rev.} D}
\def\ZPC{{\em Z. Phys.} C}
\def\JPG{{\em J. Phys.} G}
\def\EPJ{{\em Eur. Phys. J.} C}
\def\RPP{{\em Rep. Prog. Phys.}}
\title{$\pi^{\pm}~$ Spectra in Au+Au Collisions at \srt=62.4 GeV}
\author{Zhangbu Xu$^{1}$ for the STAR Collaboration}
\address{$^{1}$Physics Department, Brookhaven National Laboratory, 
Upton, NY 11973, USA}
\maketitle
\begin{abstract}
The combination of the ionization energy loss (dE/dx) from Time
Projection Chamber (TPC) at $\simeq8$\% resolution and multi-gap
resistive plate chamber time-of-flight (TOF) at 85$ps$ provides
powerful particle identification. We present spectra of identified
charged pions from transverse momentum $p_T\simeq0.2$ GeV/c to ~7-8
GeV/c in Au+Au collisions at $\sqrt{s_{_{NN}}}=62.4$ GeV. Physics
implications will be discussed.
\end{abstract}
  In this report, we studied charged $\pi^{\pm}$ spectra and their
nuclear modification factors in Au+Au collisions at \srt=62.4 GeV.
Identification of $\pi^{\pm}$ was made possible by ionization energy
loss (dE/dx) of charged particle traversing Time Projection Chamber
(TPC)~\cite{tpc}, and time-of-flight (TOF) from a prototype multi-gap
resistive plate chamber~\cite{mrpctof}. The relativistic rise of dE/dx
({\bf rdE/dx}) in TPC alone can identify $\pi^{\pm}$ from
$3^{<}_{\sim}p_T{}^{<}_{\sim}10$ GeV/c. TOF extends the identification
to lower \pto, and help confirm the dE/dx calibration.

Ionization energy loss of charged particles traversing TPC has been
used to identify hadrons at low momentum in STAR for identified
particle spectra, flow, and HBT studies.  The relativistic rise of
dE/dx also separates the electrons from the hadrons. With additional
hadron rejection capability from TOF, we were able to develop a
hadron-blind detector~\cite{starcharm} for lepton identification at
low \pto. The measurement of dE/dx is a truncated-mean method in which
30\% at the high tail of dE/dx samples in a track are
discarded~\cite{tpc}. The dE/dx resolution has been found to be
between 6\% and 11\%, depending on the magnetic field setting, event
multiplicity, beam luminosity, track length and drift distance. It was
calibrated to be better than 8\% in Au+Au collisions at \srt=62.4 GeV
for tracks with 70 cm in length. At momentum $p^{>}_{\sim}3$ \GeVc,
the dE/dx of $\pi^{\pm}$ has a $\sim15\%$ ($\sim2\sigma$) separation
from that of $K^{\pm}$ and $p(\bar{p})$ due to the relativistic rise
of pion dE/dx at large $\beta\gamma$. This allows us to identify
charged pions at $3^{<}_{\sim}p_T{}^{<}_{\sim}10$ GeV/c.
\begin{figure}[h]
  \begin{minipage}{2.45in}
    {\includegraphics[width=1.0\textwidth] {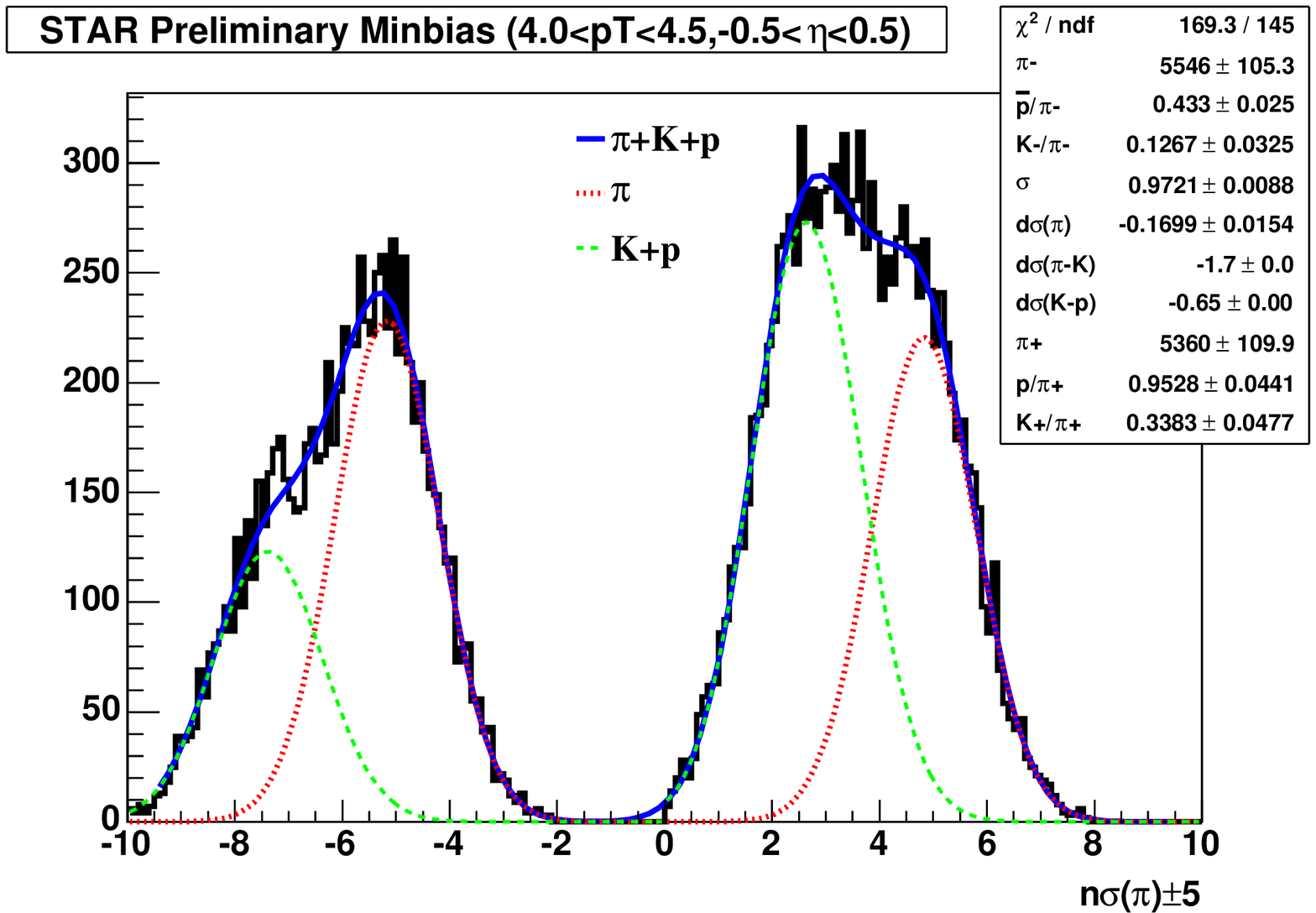}}
    \caption[]{dE/dx distribution normalized by pion
      dE/dx and offset by $\pm5$ for positive and negative charge at
      $4<p_T<4.5$ GeV/c, respectively. The distribution is from minimum-bias
      Au+Au collisions.}
    \label{fig1pid}
  \end{minipage}
  \begin{minipage}{2.45in}
    {\includegraphics[width=1.0\textwidth] {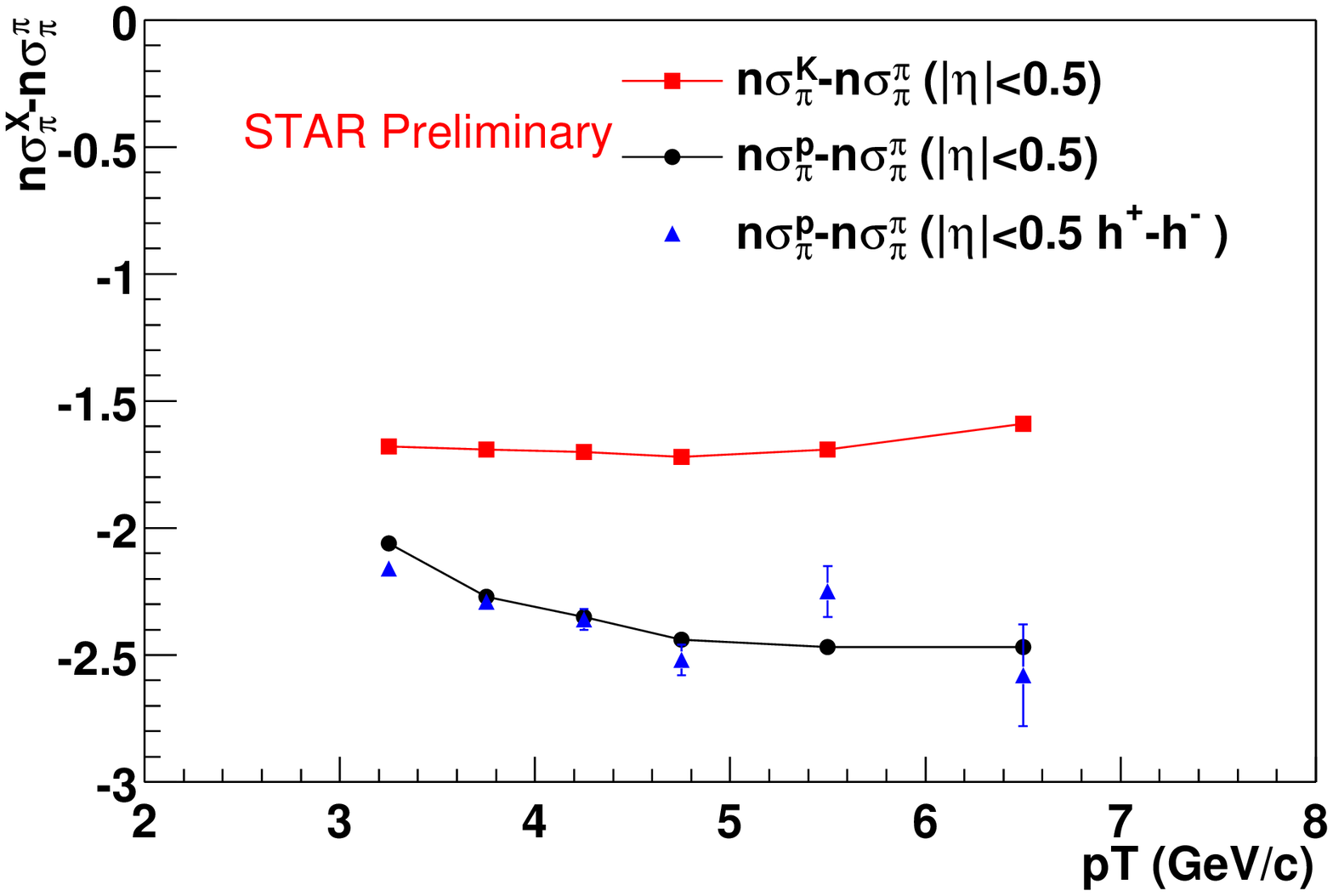}}
    \caption[]{The relative dE/dx peak position of
      K-$\pi$ (squares) and p-$\pi$ (circles) in unit of standard resolution
      width ($\sigma_{\pi}$) of pion dE/dx as function of \pto.  The
      triangles are the peak positions of the dE/dx distribution of
      $(h^+-h^{-})$.}
    \label{fig2diff}
  \end{minipage}
\end{figure}
 
The dE/dx measurement presented in Fig.~\ref{fig1pid} uses a
normalized dE/dx:
$n\sigma^{_{Y}}_{_{X}}=\ln{((dE/dx)^{_{Y}}/I_{0_{X}})}/\sigma_{_{X}}$
where $X,Y$ can be $e^{\pm},\pi^{\pm},K^{\pm}$ or $p(\bar{p})$, and
$I_{X0}$ is the expected dE/dx of particle $X$. With perfect
calibration, the $n\sigma_{\pi}^{\pi}$ distribution will be a normal
Gaussian distribution, and $n\sigma_{\pi}^{p}$ at high \pt will be a
Gaussian peaking at negative value due to smaller dE/dx of proton
traversing TPC.  Fig.~\ref{fig2diff} shows \pt dependence of
$n\sigma^{K}_{\pi}$ and $n\sigma^{p}_{\pi}$ as predicted by Bichsel
Functions for the energy loss in thin layers of argon~\cite{tpc}.  As
shown in Fig.~\ref{fig1pid} and measurements from
TOF~\cite{ming62hq04},
$(h^+-h^{-})=(p-\bar{p})+(K^+-K^-)+(\pi^+-\pi^-)\simeq
p-\bar{p}$. Therefore, the peak positions of dE/dx distribution from
$(h^+-h^{-})$ should represent well that of proton. Indeed, the
calibrated Bichsel Function for proton matches well with the dE/dx
peak position of $(h^+-h^{-})$. In addition, the dE/dx difference of
protons and pions in the momentum region where PID selection is
possible by TOF has been checked to be consistent with Bichsel
Function.  These give us confidence that Bichsel Function can be used
to constrain the relative dE/dx position between kaons, protons and
pions. In the following analysis, we fixed
$n\sigma^{K}_{\pi}-n\sigma^{\pi}_{\pi}$ and
$n\sigma^{p}_{\pi}-n\sigma^{\pi}_{\pi}$ in a six Gaussian fit, where
the six Gaussians are for $\pi^{\pm},K^{\pm}$ and $p(\bar{p})$ at a
given \pt bin.

\begin{figure}[h]
\begin{minipage}{2.45in}
{\includegraphics[width=1.0\textwidth] {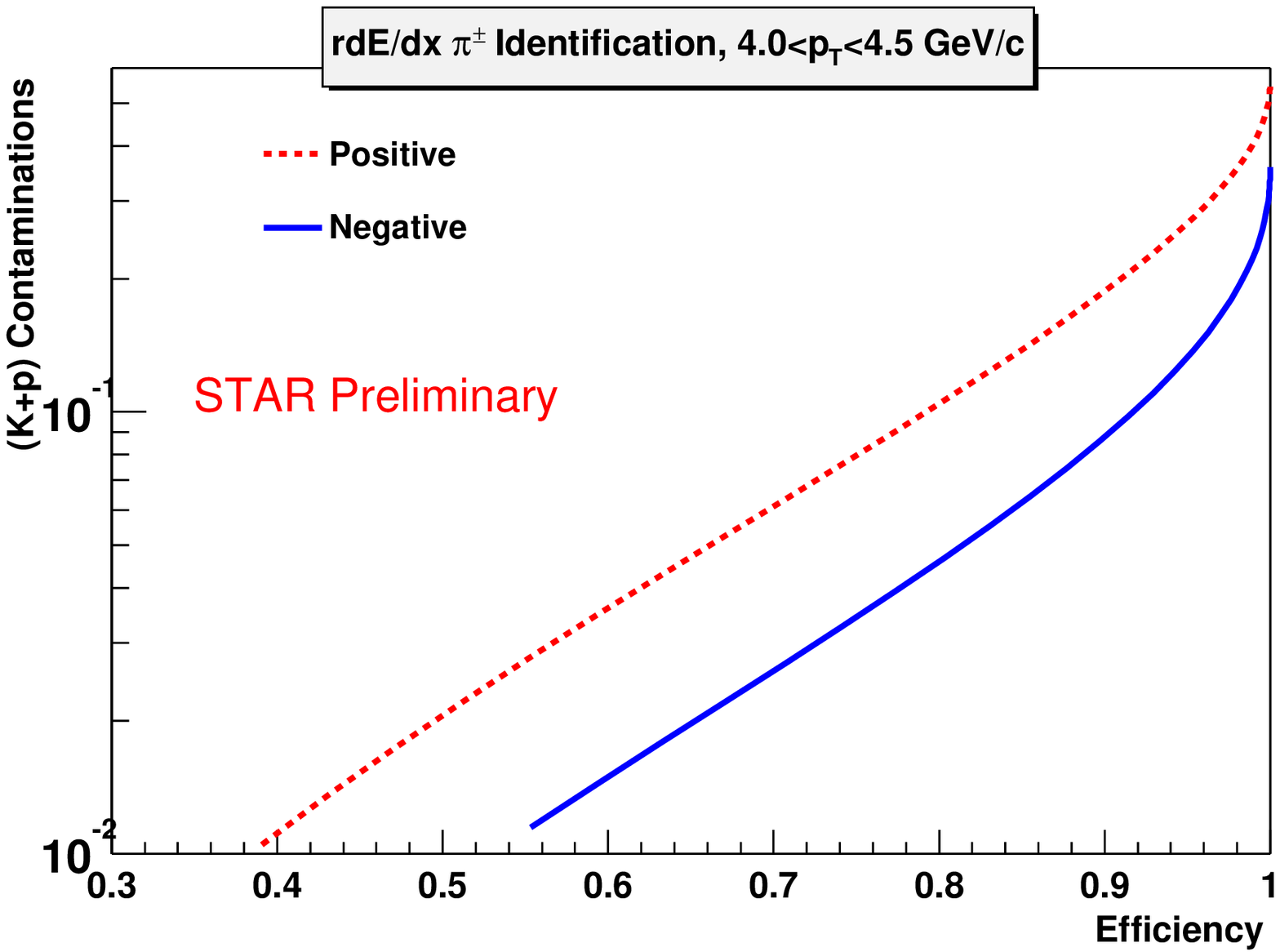}}
 \caption[]{Contamination from K+p to pions with a dE/dx selection
window vs pion efficiency from the same selection. Dashed line is for
$\pi^+$ and solid line is from $\pi^-$.}
 \label{fig3contam}
\end{minipage}
\begin{minipage}{2.45in}
{\includegraphics[width=1.0\textwidth] {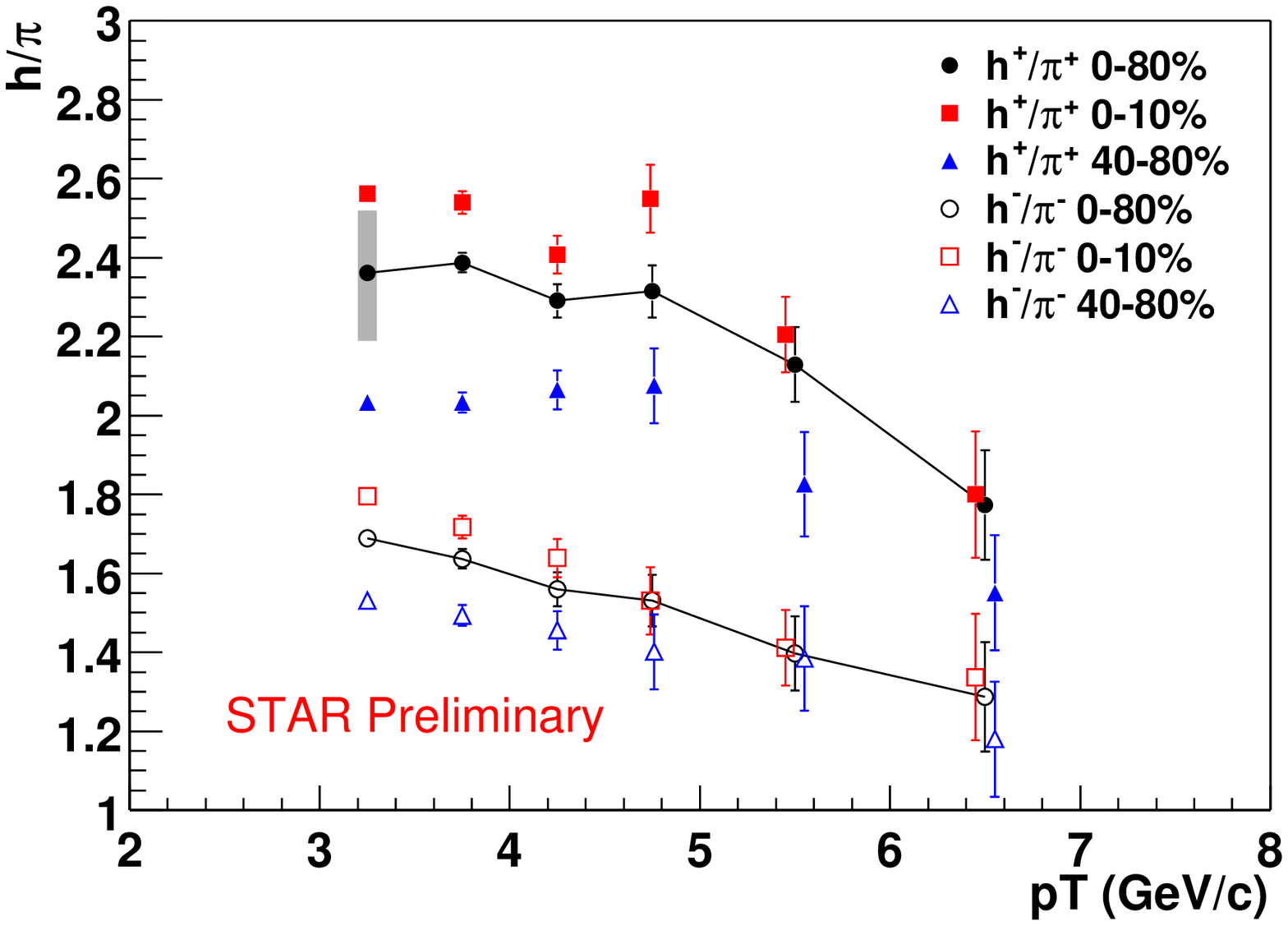}}
 \caption[]{$h^{\pm}/\pi^{\pm}$ ratios vs \pt for three centralities.
We varied $n\sigma^{K,p}_{\pi}$ around Bichsel Function to estimate
the systematical uncertainty which is highly correlated among
centralities. }
 \label{fig4h2pi}
\end{minipage}
\end{figure}
Although a fit procedure was used to extract the particle yields, pion
identification can be done track by track instead of
statistically. Fig~\ref{fig3contam} illustrates the contamination from
kaons and protons to pions when we select $n\sigma_{\pi}$ greater than
a given value. We can achieve $>95\%$ pion purity at 50\% pion
efficiency with a cut of $n\sigma_{\pi}>0$. As shown in
Fig.~\ref{fig2diff}, the separation of dE/dx of protons from kaons is
close to $1\sigma$ at high momentum, studies are underway to reliably
identify protons.  These open doors to identified particle jet
spectra, correlation, anisotropic flow at high \pt with high
statistics.

\begin{figure}[h]
\begin{minipage}{2.45in}
{\includegraphics[width=1.0\textwidth] {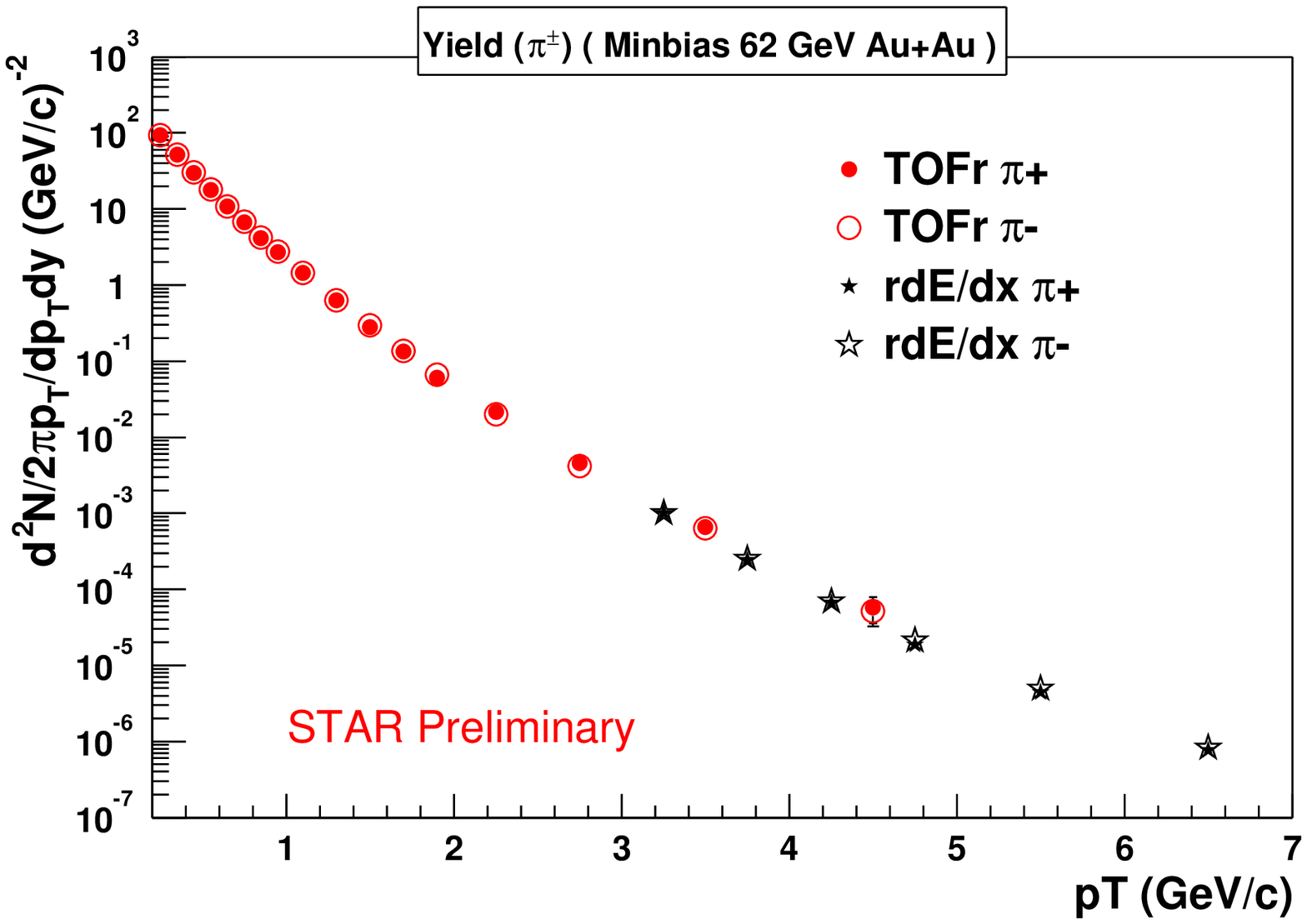}}
 \caption[]{$\pi^{\pm}$ spectra in minimum-bias Au+Au collisions from
 TOF and TPC rdE/dx. $\pi^+/\pi^-\simeq1$ while $(h^+/h^-)$ is much
 larger than unity.}
 \label{fig5spectra}
\end{minipage}
\begin{minipage}{2.45in}
{\includegraphics[width=1.0\textwidth]
{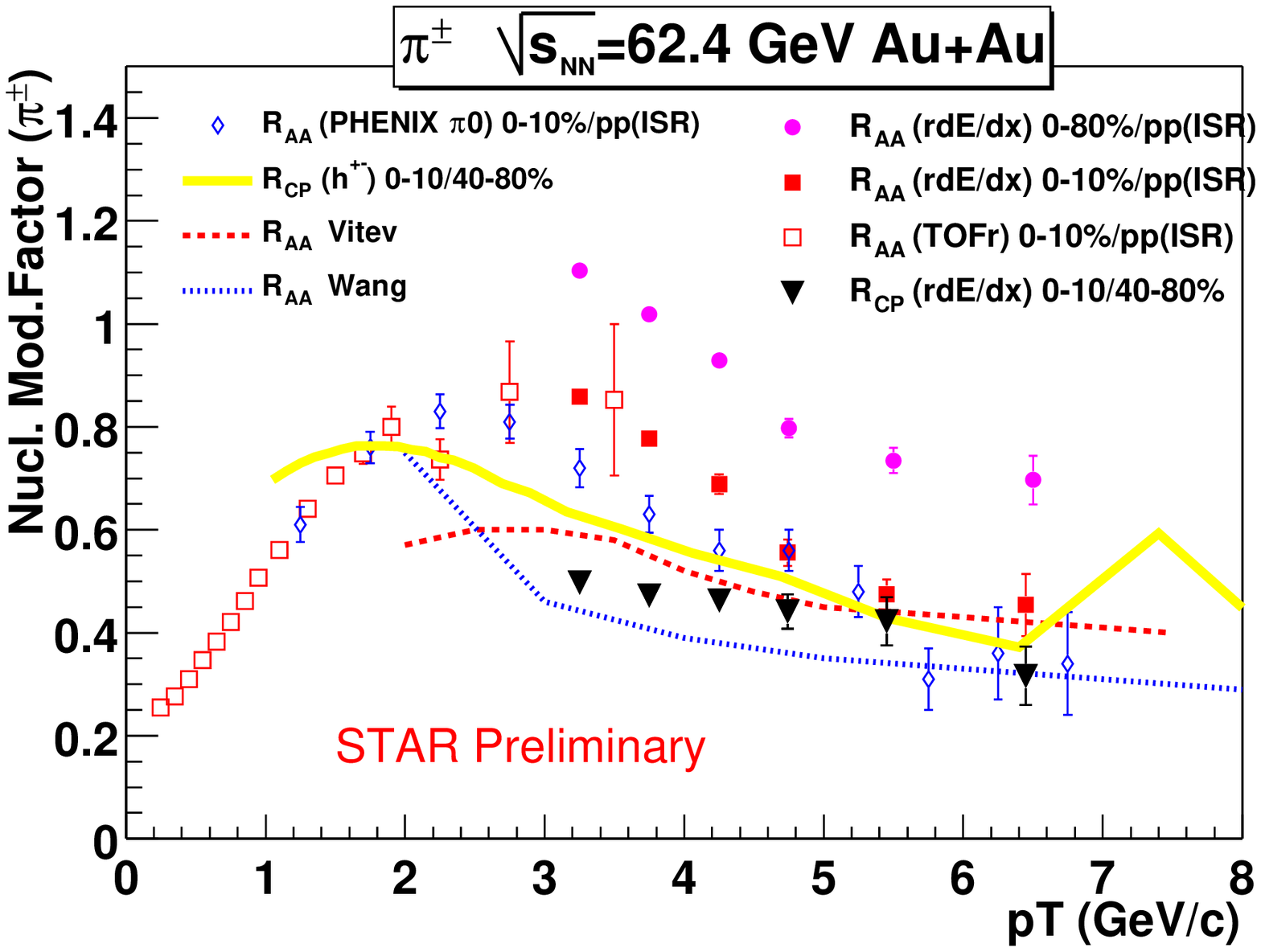}}
 \caption[]{Nuclear modification factors ($R_{AA}$ and $R_{CP}$) for
$\pi^{\pm,0}$ and inclusive hadrons. Reference spectrum of p+p is from
ISR~\cite{phenix62hq04}.}
 \label{fig6raa}
\end{minipage}
\end{figure}
Since $\pi^{\pm}$ and inclusive hadrons ($h^{\pm}$) can be identified
in the same detector with only small dE/dx difference, we can obtain
$h^{\pm}/\pi^{\pm}$ with high precision. Fig.~\ref{fig4h2pi} shows
$h^{\pm}/\pi^{\pm}$ ratios vs \pt for three centralities. The
$h^{\pm}/\pi^{\pm}$ in central collisions approach those in peripheral
collisions at $p_T{}^{>}_{\sim}5$ GeV/c, indicating an absence of
particle type dependence of nuclear effect at high $p_T$. Difference
of particle composition between negative and positive hadrons is
evident.  Fig.~\ref{fig5spectra} shows $\pi^{\pm}$ spectra in
minimum-bias Au+Au collisions from TOF~\cite{ming62hq04} and TPC
rdE/dx. The two spectra agree well within statistical
uncertainty. Fig.~\ref{fig6raa} shows nuclear modification factors of
charged pions, and the comparisons to those of
$\pi^{0}$~\cite{phenix62hq04}, inclusive hadrons~\cite{agsrhicusers}
and theoretical predictions~\cite{vitev62,wang62}. At
$p_T{}^{>}_{\sim}5$ GeV/c, $R_{AA}$ and $R_{CP}$ approach each other
and agree with the predictions. At lower $p_T$, $R_{AA}$, $R_{CP}$ and
the predictions diverge.


\begin{thebibliography}{0}
\bibitem{vitev62} I. Vitev, Arxiv: nucl-th/0404052, Curve shown is
  calculation with $dN_{g}/dy=650$.
\end{thebibliography}
\end{document}